\documentclass[aps,twocolumn,superscriptaddress,altaffilletter,lengthcheck, tightenlines,showpacs,showkeys]{revtex4}

\newcommand{\al}{\alpha}
\newcommand{\D}{\Delta}
\newcommand{\ben}{\begin{eqnarray}}
\newcommand{\een}{\end{eqnarray}}
\newcommand{\be}{\begin{equation}}
\newcommand{\ee}{\end{equation}}
\newcommand{\ba}{\begin{eqnarray}}
\newcommand{\ea}{\end{eqnarray}}
\newcommand{\n}{\label}
\newcommand{\no}{\noindent}
\newcommand{\la}{\lambda}
\newcommand{\La}{\Lambda}
\newcommand{\ga}{\gamma}
\newcommand{\ro}{\rho}

\newcommand{\bn}{\begin{equation}\label}

\usepackage{color}

\begin{document}

\title{Exactly solved models of interacting dark matter and dark energy}

\author{Luis P. Chimento}\email{chimento@df.uba.ar}

\affiliation{Departamento de F\'\i sica, Facultad
de Ciencias Exactas y Naturales, Universidad de Buenos Aires,
Ciudad Universitaria, Pabell\'on I, 1428 Buenos Aires,
Argentina,}

\bibliographystyle{plain}

\begin{abstract}
We introduce an effective one-fluid description of the interacting dark sector in a spatially flat Friedmann-Robertson-Walker space-time and investigate the stability of the power-law solutions. We find the ``source equation" for the total energy density and determine the energy density of each dark component. We study linear and nonlinear interactions which depend on the dark matter and dark energy densities, their first derivatives, the total energy density with its derivatives up to second order and the scale factor. We solve the evolution equations of the dark components for both interactions, examine exhaustively several examples and show cases where the problem of the coincidence is alleviated. We show that a generic nonlinear interaction gives rise to the ``relaxed Chaplygin gas model" whose effective equation of state includes the variable modified Chaplygin gas model while some others nonlinear interactions yield de Sitter and power-law scenarios.

\end{abstract}
\vskip 1cm


\date{\today}
\maketitle

\section{Introduction}

At large scales, there are strong evidences for a spatially flat and accelerating universe  transiting from a scenario dominated by matter accumulated by purely attractive gravitational effects to another dominated by a dark energy component dispersed by repellent gravitational effects and characterized by a negative pressure \cite{Sahni:1999gb}. This behavior has lasted till present and most likely  continue for ever. The reasons, causes and details of when this transition happened, it was not still understood. For a sufficiently intense acceleration one can speak rather of superacceleration, then the possibility exists that the universe has a catastrophic end with a suddenly future singularity at a finite time (Big Rip) and a total disintegration of the well-known structures. 

Astronomical observations suggest that the Universe entered an accelerated expansion stage when the value of its scale factor was approximately one-half of the current one. This important discovery, which was based on the observations of the brightness of a class of supernovae (SNIa) \cite{Riess:1998cb}, has been confirmed by precise measurements of the spectrum of the cosmic microwave background (CMB) anisotropies \cite{Bennett} as well as the baryon acoustic oscillations in the Sloan digital sky survey luminous galaxy sample \cite{Eisenstein:2005su}. The consensus between cosmologist points in the direction that the understanding of the phenomenon will probably require a unified comprehension of the gravitational and the other fundamental interactions.

We will consider fundamentally interacting and unified models to investigate the mechanisms governing the dynamics of the evolution of the universe from its early stage until its recent accelerated phase.

In the interacting models, the source of Einstein equations includes an aggregate of different material fluids that interact among them. This is in principle the simplest, and perhaps the most obvious hypothesis, and it is in fact, the one that has provided more advances in the knowledge of the phenomenon of the recent acceleration of the universe \cite{Binder:2006yc}, \cite{waga}, and references therein. Observational evidences show three fundamental components: baryons, dark matter and dark energy. Given the dynamical similarity between baryons and dark matter, it is possible to make a simplified model replacing both components with a nearly pressureless dust. 

In unified models, the Einstein equations will have a single component working as dark matter and dark energy at different stages. It interpolates smoothly between a matter dominated phase in the early stage and dark energy in the late stage of the evolution, so inducing an accelerated expansion of the universe. The Chaplygin gas and its extensions were the unified models that have been more studied in the literature \cite{Kamenshchik}-\cite{Cervantes-Cota}. More recently, several generalizations of those models have been introduced, as for instance, the variable modified Chaplygin gas model with an equation of state, depending explicitly on the scale factor \cite{guo1}-\cite{deb}. 

In the following we will use either interacting or unified models indistinctly because there is some sort of resemblance between these different models.
  
In section II we consider a dark sector model with energy transfer, develop an effective one-fluid description, find the source equation for the total energy density and investigate the stability of the power-law solutions. In section III we introduce the ``linear interaction", impose the stability condition, describe some simple interacting models and find the exact scale factor as well as the effective barotropic index. Also, we present a ``general linear interaction" which induces a generalized $\La$CDM model. In section IV we investigate a ``nonlinear interaction", solve the source equation and show that the effective equation of state of the dark sector includes several generalizations of the Chaplygin gas. Finally, in section V the conclusions are stated.

%%%%%%%%%%%%%%%%%%%%%%%%%%%%%%%%%%%%%%%%%%%%%%%%%%%%%% 
\section{Dark sector evolution}
%%%%%%%%%%%%%%%%%%%%%%%%%%%%%%%%%%%%%%%%%%%%%%%%%%%%%%

%%%%%%%%%%%%%%%%%%%%%%%%%%%%%%%%%%%%%%%%%%%%%%%%%%%%%% 
\subsection{Effective one-fluid description}
%%%%%%%%%%%%%%%%%%%%%%%%%%%%%%%%%%%%%%%%%%%%%%%%%%%%%%

We consider an expanding universe modeled by a mixture of interacting dark matter and dark energy with energy densities $\ro_c$ and $\ro_x$, and pressures $p_c$ and $p_x$ respectively. Due to the energy transfer between both dark components, they do not evolve separately \cite{Binder:2006yc} and the Einstein equations read
\be
\label{00i}
3H^{2}=\ro_c+\ro_x,
\ee
\be
\n{coi}
\dot\ro_c+\dot\ro_x+3H(\ro_c+p_c+\ro_x+p_x)=0.
\ee

\no where $a$ is the scale factor and $H=\dot a/a$. The conservation equation (\ref{coi}) evidences the interaction among the components admitting the mutual exchange of energy and momentum. 

For the two dark components we assume  equations of state $p_c=(\ga_c-1)\ro_c$ and $p_x=(\ga_x-1)\ro_x$, where the barotropic indices $\ga_c$ and $\ga_x$ are constants. The dark matter is composed of nearly pressureless components with a barotropic index $\ga_c\approx 1$ and the dark energy violates the strong energy condition, $\ro_x+3p_x<0$, so that $\ga_x<2/3<\ga_c$. The total energy density $\ro$ and the conservation equation for the interacting two-fluid model are
\be
\n{c1}
\ro=\ro_c+\ro_x,
\ee
\be 
\n{c2}
\ro'=-\ga_c\ro_c-\ga_x\ro_x,
\ee
where the prime indicates differentiation with respect to the new time variable $'\equiv d/d\eta=d/3Hdt=d/d\ln{(a/a_0)^3}$ and $a_0$ is some value of reference for the scale factor. Solving the system of  equations (\ref{c1})-(\ref{c2}) we get $\ro_c$ and $\ro_x$ as functions of $\ro$ and its derivative $\ro'$ 
\be
\n{31}
\ro_c=-\frac{\ga_x\ro+\ro'}{\D\,}, \qquad \ro_x=\frac{\ga_c\ro+\ro'}{\D},
\ee
where $\D=\ga_c-\ga_x>0$ is the determinant of the linear equation system (\ref{c1})-(\ref{c2}).  

At this point, we introduce an energy transfer between the two fluids by separating the conservation equation (\ref{c2}) into the two equations  
\be
\n{de1}
\ro'_c+\ga_c\ro_c=-Q,
\ee\be
\n{de2}
\ro'_x+\ga_x\ro_x=Q.
\ee
Here, we have consider a coupling with a factorized $H$ dependence $3HQ$ where the interaction term $Q$ generates the energy transfer between the two fluids. So the dynamics of $\ro_c$ and $\ro_x$ is dictated by the scale factor instead of $H$.  Differentiating the first or the second Eq. (\ref{31}) and combining with the Eq. (\ref{de1}) or with the Eq. (\ref{de2}), we obtain a second order differential equation for the total energy density  
\be
\n{2} 
\rho''+(\ga_c+\ga_x)\rho'+\ga_c\ga_x\rho= Q\D,
\ee
that we call the "source equation" \cite{luis}.

The interacting two-fluid model has been reduced to an effective one-fluid model with total energy density $\ro$ and total pressure $p=p_c+p_x$, whose effective equation of state is
\be
\n{pe}
p(\ro,\ro')=-\ro-\ro'.
\ee 
This enable us to assume an effective one-fluid description with equation of state $p=(\gamma-1)\rho$ and effective conservation equation $\ro'+\ga\ro=0$, where the effective barotropic index $\gamma=(\ga_c\ro_c+\ga_x\ro_x)/\rho$ ranges between $\ga_x<\ga<\ga_c$. 

Resuming, given an interaction $Q$, the total energy density $\ro$ of the effective one-fluid model is determined by  solving the source equation Eq. (\ref{2}). Once we know $\ro$, we are able to find the effective equation of state from Eq. (\ref{pe}) and the scale factor by integrating the Friedmann equation $3H^2=\rho$, without knowing $\ro_c$ and $\ro_x$ separately. Both energy densities are easily calculated by replacing $\ro$ and $\ro'$ into the Eq. (\ref{31}). For instance, in the no interaction case, $Q=0$, the  energy density of the effective one-fluid model is $\ro=c\,/a^{3\ga_c}+b\,/a^{3\ga_x}$. For any value of the constants $c$ and $b$,  $\ro\to b\,/a^{3\ga_x}$, the scale factor $a\to t^{2/3\ga_x}$ and the power-law solution  $t^{2/3\ga_x}$ becomes an attractor. Throughout the paper $c_i$, $b_i$, $j_i$, $k_i$ and $n_i$ with $i=1,2,3....$, will represent constants while $c$ and $b$ will be integration constants.

Basically, we have shown that an interacting two-fluid model can be seen as an effective one-fluid model or equivalently considered as a unified one. Its dynamics is given by the two independent Einstein equations 
\be
\label{00}
3H^{2}=\rho, \qquad
\dot\rho+3H(\rho+p)=0.
\ee
These equations cannot determine the three quantities $a$, $p,$ and $\rho $ because there is one degree of freedom. Usually, the system of equations (\ref{00}) is closed with an equation of state $p=p(\rho)$. On the other hand, when we split the effective energy-momentum tensor into two dark components, the Eqs. (\ref{00}) become Eqs. (\ref{00i}) and (\ref{coi}), which cannot determine the five quantities $a$, $\ro_c$, $\ro_x$, $p_c$ and $p_x$. Thus, we need to introduce an equation of state for each dark component $p_c=(\gamma_c-1)\rho_c$ and $p_x=(\gamma_x-1)\rho_x$ to preserve the one degree of freedom of the unified model. Then, by changing these equations of state we obtain a very large set of interacting models which are equivalent to a unified one, meaning that the decomposition into dark  matter and dark energy is not unique. 

%%%%%%%%%%%%%%%%%%%%%%%%%%%%%%%%%%%%%%%%%%%%%%%%%%%%%%%%
\subsection{Asymptotic stability}
%%%%%%%%%%%%%%%%%%%%%%%%%%%%%%%%%%%%%%%%%%%%%%%%%%%%%%%%

The knowledge of stable power-law solutions $a=t^{2/3\ga_s}$ for the interacting two-fluid model is very useful because these attractor solutions determine the asymptotic behavior of the effective barotropic index $\ga=-2\dot H/3H^2=\ga_s$. These solutions are associated to a universe approaching to a stationary stage i.e., existence of attractor solutions $\ga_s$. Consistently $\ga$ tends to the asymptotic constant value $\ga_s$. Then on the attractor
\be
\n{or}
\ga_s=\frac{\ga_c\ro_{cs}+\ga_x\ro_{xs}}{\ro_{cs}+\ro_{xs}}=\frac{r_s\ga_c+\ga_x}{1+r_s},
\ee
we conclude that the ratio $r=\ro_c/\ro_x$ turns asymptotically constant $r_s=\ro_{cs}/\ro_{xs}$, alleviating the problem of the coincidence: are the proportions of matter accumulated by gravitative effects and dark energy comparable at the present time for a strange coincidence or for a fundamental reason? \cite{luis}-\cite{Jackson}. For the case $\ga_s=0$, we have a final de Sitter stage, $H= const$, with $r_s=-\ga_x/\ga_c$. 

To investigate the stability of the constant solution $\ga_s$ we use the evolution equation of $\ga$, which is obtained by replacing $\ro'=-\ga\ro$ and $\ro''=(\ga^2-\ga')\ro$ into the source equation (\ref{2}), so we have
\be
\n{41}
\ga'-(\ga-\ga_c)(\ga-\ga_x)=-\frac{\D}{\ro}\,\,Q.
\ee

We proceed assuming that a constant solution $\ga=\ga_s$ of the Eq. (\ref{41}) with $\ga_x<\ga_s<\ga_c$  exists, and after that, we will impose the stability condition so that $\ga_s$ be stable. An interaction satisfying the existence requirement belongs to the class 
\be
\n{qe}
Q(\ga_s)=\D^{-1}(\ga_s-\ga_c)(\ga_s-\ga_x)\,\ro,
\ee
with $Q(\ga_s)<0$. The negative value of $Q(\ga_s)$ on the attractor indicates that the energy is being transferred from dark energy to dark matter. It assures that power-law solutions $a=t^{2/3\ga_s}$, obtained after integrate $\ga_s=-2\dot H/3H^2$, are stable and the ratio $r$ tends asymptotically to the constant value $r_s$. 

The analysis of stability will be performed for interactions that have the form $Q=Q(\ro_c,\ro_x,\ro_c',\ro_x',\ro,\ro',\ro'')$. By using the Eq. (\ref{31}) and $\ro'=-\ga\ro$ with $\ro''=(\ga^2-\ga')\ro$, we obtain that $\ro_{c,x}=\ro_{c,x}(\ro,\ro')$ and $\ro_{c,x}'=\ro_{c,x}'(\ro',\ro'')$, then the interaction becomes $Q=Q(\ga,\ga',\ro)$. For simplicity we  adopt separability of $Q$, that is, $Q=Q(\ga,\ga',\ro)=\ro \,Q(\ga,\ga')$ and write 
\be
\n{42}
Q(\ga,\ga',\ro)=\D^{-1}(\ga-\ga_c)(\ga-\ga_x)\,F(\ga,\ga')\ro ,
\ee
where the function $F$ depends on $\ga$ and $\ga'$. Several interacting models analyzed in the literature are described by the interaction term (\ref{42}), see for instance \cite{gilberto}-\cite{Barrow}. Other kinds of interactions were explored in \cite{alcaniz}. 

By combining Eqs. (\ref{41}) and (\ref{42}) we rewrite the evolution equation of the effective barotropic index as  
\be
\n{gaf}
\ga'=-(\ga-\ga_c)(\ga-\ga_x)(F-1).
\ee 
Hence, when the function $F$ fulfills the two conditions 
\be
\n{43}
F(\ga=\ga_s,\ga'=0)=1,
\ee
and 
\be
\n{cs}
\left(\frac{\partial\ga'}{\partial\ga}\right)_{(\ga_s,0)}=-\frac{(\ga_s-\ga_c)(\ga_s-\ga_x)F_{\ga}(\ga_s,0)}{1+(\ga_s-\ga_c)(\ga_s-\ga_x)F_{\ga'}(\ga_s,0)}<0,
\ee
where $F_{\ga}$ and $F_{\ga'}$ stand for the partial derivatives of $F$ with respect to $\ga$ and $\ga'$ respectively, then $\ga_s$ is a stable solution or an attractor. In other words, when the condition (\ref{43}) is satisfied $\ga_s$ becomes a constant stationary solution of the  Eq. (\ref{gaf}) and it is stable whenever the stability condition (\ref{cs}) is fulfilled. 

%%%%%%%%%%%%%%%%%%%%%%%%%%%%%%%%%%%%%%%%%%%%%%%%%%%%%
\section{Linear interaction}
%%%%%%%%%%%%%%%%%%%%%%%%%%%%%%%%%%%%%%%%%%%%%%%%%%%%%

Interaction terms depending linearly on the energy densities $\ro_c$, $\ro_x$ and the total energy density $\ro$ have been investigated in a broad class of cosmological models  \cite{luis}-\cite{Valiviita}. When the dark matter and dark energy are coupled with $\ro_x$, $\ro$ or linear combinations of $\ro_c$ and $\ro_x$, the ratio $r$ may tend asymptotically to a constant value. This attractive feature encourage us to explore an extended set of interactions. From Eq. (\ref{31}), we have seen that $\ro_c$ and $\ro_x$ are linear functions of $\ro$ and its derivative $\ro'$. Taking into account that the latter term includes the first derivative of the energy densities $\ro_c'$, $\ro_x'$, and 
from Eq. (\ref{31}), these terms introduce a dependence with $\ro''$, we will proceed to study the case of considering the ``linear interaction" $(Q_L)$ which admits new terms proportionals to $\ro_c'$, $\ro_x'$ and $\ro''$. Then, we start building the ``linear interaction", $Q_L$, as the following linear combination
\be
\n{lc}
Q_L=c_1\ro_c+c_2\ro_x+c_3\ro_c'+c_4\ro_x'+c_5\ro+c_6\ro'+c_7\ro'',
\ee 
in such a way that it verifies the conditions (\ref{43})-(\ref{cs}). The  interaction $Q_L$ is also motivated for the fact that the source equation (\ref{2}) becomes a linear second order differential equation for $\ro$. As far as we know, this this kind of coupling was not investigated in the literature, so we will analyze in detail the $Q_L$. The particular cases $Q_L=c_1\ro_c+c_2\ro_x$ were previously investigated in Refs. \cite{Sadjadi}, \cite{Barrow}  with $Q_L=c_5\ro$ in \cite{zpc}-\cite{Wang}, with $Q_L=c_1\ro_c$ in \cite{Amendola} and with $Q_L=c_2\ro_x$ in \cite{gilberto}-\cite{Jackson}. 

By using the Eqs. (\ref{31}) in the Eq. (\ref{lc}), the $Q_L$ can be reduced to a linear combination of the basis elements $\ro$, $\ro'$ and $\ro''$. Finally we get
\be
\n{lcg}
Q_L=b_1\ro+b_2\ro'+b_3\ro'',
\ee 
where the constants $b_i$ are linear combinations of the constants $c_i$ in the $Q_L$ (\ref{lc}),
\be
\n{c1'}
b_1=\D^{-1}(-\ga_x c_1+\ga_c c_2+c_5),
\ee
\be
\n{c2'}
b_2=\D^{-1}(-c_1+c_2-\ga_x c_3+\ga_c c_4+c_6),
\ee
\be
\n{c3}
b_3=\D^{-1}(-c_3+c_4+c_7).
\ee
Combining $\ro'=-\ga\ro$ and $\ro''=(\ga^2-\ga')\ro$ with the Eq. (\ref{lcg}), and comparing the $Q_L$ (\ref{lcg}) with the Eq. (\ref{42}) we obtain the following function $F(\ga,\ga')$ 
\be
\n{fgl}
F=\frac{\D}{(\ga-\ga_c)(\ga-\ga_x)}\left[b_1-b_2\ga+b_3(\ga^2-\ga')\right].
\ee
The existence of a stationary solution $\ga_s$ of the Eq. (\ref{gaf}) is linked to that the function $F$ satisfies the condition (\ref{43}) leading to the constrain 
\be
\n{bc'}
\frac{\D}{(\ga_s-\ga_c)(\ga_s-\ga_x)}\left[b_1-b_2\ga_s+b_3\ga_s^2\right]=1,
\ee
for the coefficient $b_1$, $b_2$ and $b_3$. By solving Eqs. (\ref{bc'}) for $b_2$ and inserting it into Eq. (\ref{lcg}), we obtain the final form of the $Q_L$ 
\be
\n{qlf}
Q_L=b_1\ro+\ga_s^{-1}\left[b_1+b_3\ga_s^2-\frac{(\ga_s-\ga_c)(\ga_s-\ga_x)}{\D}\right]\ro'+b_3\ro''.
\ee
By inserting this $Q_L$ into the Eq. (\ref{41}) we find the two constant solutions
\be
\n{17}
\ga^-=\ga_s,    \qquad   \ga^+=\frac{\ga_c\ga_x-b_1\D}{\ga_s(1-b_3\D)},
\ee
while the stability condition (\ref{cs}) gives
\be
\n{18}
\ga_s-\ga^+<0,
\ee
so, the constant solution $\ga_s$ is asymptotically stable provided that  $\ga_s<\ga^+$ with the additional requirement $\ga_x<\ga_s<\ga^+<\ga_c$, yielding a stable cosmological model with the  asymptotic power-law expansion $a=t^{2/3\ga_s}$.

The exact general solution of the source equation (\ref{2}) for $Q_L$ (\ref{qlf}) and the effective equation of state (\ref{pe}) are 
\be
\n{sQL}
\ro_L=c\,a^{-3\ga_s}+b\,a^{-3\ga^+},
\ee
\be
\n{p12}
p_L=(\ga_s-1)\ro_L+(\ga^+-\ga_s)b\,a^{-3\ga^+}.
\ee
The interacting model is finally realized when the general solution (\ref{sQL}) is inserted into the energy density of each dark component (\ref{31})
\be
\n{r12'}
\ro_{c}=\frac{\left(\ga_s-\ga_x\right)c\,a^{-3\ga_s}+\left(\ga^+-\ga_x\right)b\,a^{-3\ga^+}}{\D},
\ee
\be
\n{r22}
\ro_{x}=\frac{\left(\ga_c-\ga_s\right)c\,a^{-3\ga_s}+\left(\ga_c-\ga^+\right)b\,a^{-3\ga^+}}{\D},
\ee
For large scale factors the quantities $\ro$, $\ro_c$, $\ro_x$, $\ro_c'$, $\ro_x'$, $\ro'$, $\ro''$, $p$ and the $Q_L$ behave as  $a^{-3\ga_s}$ while in the initial regimen the above quantities behave as $a^{-3\ga^+}$.

The coupling between the two dark components modifies typical characteristics of $\ro_c$ and $\ro_x$. In fact, the universe begins with a mix of dark matter (\ref{r12'}) and dark energy (\ref{r22}) represented approximately by the unstable $\ro_c \propto (\ga^+-\ga_x)a^{-3\ga^+}$ and $\ro_x\propto (\ga_c-\ga^+)a^{-3\ga^+}$ respectively. After that, the instability of the constant solution $\ga^+$ induces the universe to evolve from that unstable era, characterized by  $r_+=(\ga^+-\ga_x)/(\ga_c-\ga^+)$, to a stable final stage where the dark matter and dark energy densities are dominated by the stable components $\ro_c\propto (\ga_s-\ga_x)a^{-3\ga_s}$ and $\ro_x\propto (\ga_c-\ga_s)a^{-3\ga_s}$. The stable solution $\ga_s$ is associated to an asymptotically  stable ratio  $r_s=(\ga_s-\ga_x)/(\ga_c-\ga_s)$ with $r_+>r_s$, showing that the linear interaction alleviates the problem of the coincidence. In turn, the scale factor interpolates between the unstable stage, evolving as  $a\propto t^{2/3\ga^+}$, and the stable stage evolving as $a\propto t^{2/3\ga_s}$. Meanwhile the effective equation of state  (\ref{p12}) plays the role of a peculiar fluid. 

%%%%%%%%%%%%%%%%%%%%%%%%%%%%%%%%%%%%%%%%%%%%%%%%%%%%%
\subsection{Linear examples}
%%%%%%%%%%%%%%%%%%%%%%%%%%%%%%%%%%%%%%%%%%%%%%%%%%%%%

We will study simple examples where the dark components interact with each other successively by considering separately only the terms $\ro_c$, $\ro_x$, $\ro$ and $\ro'$ of the $Q_L$ (\ref{lc}) and review some of the models investigated with these particular couplings. Also, we select the four function $F(\ga)$ in such a way that the condition (\ref{43}) is satisfied identically.

(1) $c_1\ne 0$
\be
\n{c_2}
Q_{\ro_c}=(\ga_s-\ga_c)\ro_c, \qquad F_{\ro_c}=\frac{\ga_s-\ga_c}{\ga-\ga_c}, 
\ee
with $Q_{\ro_c}<0$ \cite{Amendola}. The stability condition (\ref{cs}), is not satisfied and the power-law solution $a=t^{2/3\ga_s}$ becomes unstable. This model contains serious instabilities on the perturbations of the dark energy component \cite{Valiviita}.

In this case the solution of the source equation (\ref{2}) is
\be
\n{442}
\ro_{\ro_c}=c\,a^{-3\ga_s}+b\,a^{-3\ga_x}.
\ee
For any value of the initial conditions $c$,$b$ and for large scale factor, the total energy density (\ref{442}) has the limit $\ro_{\ro_c}\to b/a^{3\ga_x}$, meaning that $a\to t^{2/3\ga_x}$ since $\ga_x<\ga_s$. The model seems to be completely dominated by the input source $\ro_x$. In fact, the energy densities (\ref{31}) are
\be
\n{re2}
\ro_c=\D^{-1}\left(\ga_s-\ga_x\right)c\,a^{-3\ga_s},
\ee
\be
\n{re22}
\ro_x=\D^{-1}\left(\ga_c-\ga_s\right)c\,a^{-3\ga_s}+b\,a^{-3\ga_x},
\ee
and the ratio $r_{\ro_c}\propto a^{-3(\ga_s-\ga_x)}\to 0$. Then, at late times, the interacting two-fluid model with energy transfer $Q_{\ro_c}$ is not satisfactory. However, this coupling can work when it is combined  linearly with some of the other parts of $Q_L$.

(2) $c_2\ne 0$
\be
\n{c3'}
Q_{\ro_x}=-(\ga_s-\ga_x)\ro_x, \qquad F_{\ro_x}=\frac{\ga_s-\ga_x}{\ga-\ga_x}, 
\ee
with $Q_{\ro_x}<0$. This interaction was examined in several papers \cite{gilberto}-\cite{Jackson}. Now the stability condition (\ref{cs}) is satisfied 
and the solution $\ga_s$ is stable. By solving the source equation (\ref{2}) for $Q_{\ro_x}$, we obtain 
\be
\n{443}
\ro_{\ro_x}=c\,a^{-3\ga_s}+b\,a^{-3\ga_c}.
\ee
For any value of the initial conditions $c$,$b$ and large scale factor, the total energy density $\ro\to c_{1}/a^{3\ga_s}$ and $a\to t^{2/3\ga_s}$ because $\ga_s<\ga_c$ showing that the interacting model is dominated by the attractor $\ga_s$.

The dark matter and dark energy densities (\ref{31}) are 
\be
\n{re3}
\ro_c=\D^{-1}\left(\ga_s-\ga_x\right)c\,a^{-3\ga_s}+b\,a^{-3\ga_c},
\ee
\be
\n{re33}
\ro_x=\D^{-1}\left(\ga_c-\ga_s\right)c\,a^{-3\ga_s},
\ee
showing that the ratio $r_{\ro_x}\to r_s=(\ga_s-\ga_x)/(\ga_c-\ga_s)$ on the attractor. Then, the interaction $Q_{\ro_x}$ may represent adequately an interacting dark sector model which can be adapted to the observations. A cosmological model with the above $Q_{\ro_x}$ was proposed for the current universe which consists of noninteracting baryonic matter and interacting dark components \cite{gilberto}. The evolution of a viscous cosmology model was also analyzed by employing the energy transfer $Q_{\ro_x}$ \cite{Chen}.

(3) $c_5\ne 0$
\be
\n{c_1}
Q_\ro=\frac{(\ga_s-\ga_c)(\ga_s-\ga_x)}{\D}\,\ro,\quad F_\ro=\frac{(\ga_s-\ga_c)(\ga_s-\ga_x)}{(\ga-\ga_c)(\ga-\ga_x)}. \,\,\,\,\,\,\,\,\,
\ee
with $Q_\ro<0$. It produces a transition from a dark matter dominated phase to an accelerated expansion phase dominated by dark energy \cite{zpc}-\cite{Wang}. 

By imposing the stability condition (\ref{cs}) on the function $F_\ro$ (\ref{c_1}) we find that $\ga=\ga_s$ is an attractor provided $\ga_x<\ga_s<(\ga_c+\ga_x)/2<\ga_c$. Inserting $Q_\ro$ into the source equation (\ref{2}) we obtain the total energy density  
\be
\n{441}
\ro_\ro=c\,a^{-3\ga_s}+b\,a^{-3(\ga_c+\ga_x-\ga_s)}.
\ee
For any value of the initial conditions $c$,$b$ and large scale factor, $\ro\to c/a^{3\ga_s}$ and the power-law expansion $a\to t^{2/3\ga_s}$ becomes asymptotically stable. 

The dark matter and dark energy densities (\ref{31}) are
\be
\n{re1}
\ro_c=\frac{\left(\ga_s-\ga_x\right)c\,a^{-3\ga_s}+\left(\ga_c-\ga_s\right)b\,a^{-3(\ga_c+\ga_x-\ga_s)}}{\D},
\ee
\be
\n{re21}
\ro_x=\frac{\left(\ga_c-\ga_s\right)c\,a^{-3\ga_s}+\left(\ga_s-\ga_x\right)b\,a^{-3(\ga_c+\ga_x-\ga_s)}}{\D}.
\ee
Thus, the ratio $r_{\ro}$ tends to $r_s= (\ga_s-\ga_x)/(\ga_c-\ga_s)$, being $r_s$  an attractor.

(4) $c_6\ne 0$
\be
\n{4'}
Q_{\ro'}=\frac{(\ga_c-\ga_s)(\ga_s-\ga_x)}{\ga_s\D}\,\ro', \quad
F_{\ro'}=\frac{\ga(\ga_s-\ga_c)(\ga_s-\ga_x)}{\ga_s(\ga-\ga_c)(\ga-\ga_x)}, 
\ee 
From $\ro'=-\ga\ro<0$, we see that $Q_{\ro'}<0$ is negative. Besides, by imposing the stability condition (\ref{cs}) on the function $F_{\ro'}$, we obtain that $\ga_s$ is an attractor provided 
$\ga_s^2<\ga_c\ga_x$. Solving the source equation (\ref{2}) for $Q_{\ro'}$, we find the total energy density 
\be
\n{444}
\ro_{\ro'}=c\,a^{-3\ga_s}+b\,a^{-3\ga_c\ga_x/\ga_s}.
\ee
Then, whatever be the initial conditions $c$ and $b$ the total energy density has the limit $\ro\to c/a^{3\ga_s}$ for large scale factor, evidencing that $\ga_s$ is an attractor.

The dark matter and dark energy densities (\ref{31}) are given by
\be
\n{re4}
\ro_c=\frac{\left(\ga_s-\ga_x\right)c\,a^{-3\ga_s}+\ga_x\ga_s^{-1}\left(\ga_c-\ga_s\right)b\,a^{-3\ga_c\ga_x/\ga_s}}{\D},
\ee
\be
\n{re44}
\ro_x=\frac{\left(\ga_c-\ga_s\right)c\,a^{-3\ga_s}+\ga_c\ga_s^{-1}\left(\ga_s-\ga_x\right)b\,a^{-3\ga_c\ga_x/\ga_s}}{\D},
\ee
and $r_{\ro'}=(\ga_s-\ga_x)/(\ga_c-\ga_s)$ on the attractor. As far as we know, this interacting model was not investigated in the literature. It appears as a feasible candidate to be considered for to describe the evolution of the dark sector.

Finally, by solving the Friedmann equation $3H^2=\ro_L$ for the source (\ref{sQL}), we find the exact scalar factor 
\be
\n{a}
a_L=\left[\omega\,\sinh{\Delta\tau}\right]^{2/3(\ga^+-\ga_s)},
\ee
\be
\n{t}
t=\frac{2}{\sqrt{3b}\,(\ga^+-\ga_s)}
\int \left[\omega\sinh{\Delta\tau}\right]^{\ga_s/(\ga^+-\ga_s)}d\tau,\,\,\,\,
\ee
where $\omega^2=c/b$, see Ref. \cite{crossing}. Due to $\ga^+>\ga_s$, the latter equation shows that the variables $t$ and $\tau$ have the same asymptotic limits. Then, we use $\tau$ instead of $t$ to analyze the scale factor and the effective barotropic index  
\be
\n{ge}
\ga_L=\frac{\ga^++\ga_s\sinh^2{\omega\Delta\tau}}{\cosh^2{\omega\Delta\tau}},
\ee
in the two asymptotic regimes. So as $t$ grows the model interpolates between the initial $\ga^+$ and the final $\ga_s$ values. Eqs. (\ref{a})-(\ref{ge}) allows us to express the total energy density (\ref{sQL}), the dark matter and dark energy densities (\ref{31}), the ratio $r_L$ and the effective pressure (\ref{pe}) as functions of the new time $\tau$. In particular, at early and later times, the asymptotic limits of the ratio $r_L$ become
\be
\n{rl}
r_L^{+}=\frac{\ga^+-\ga_x}{\ga_c-\ga^+}, \qquad
r_{s}=\frac{\ga_s-\ga_x}{\ga_c-\ga_s},
\ee
which satisfy the crucial relation $r_L^{+}>r_{s}$. 

%%%%%%%%%%%%%%%%%%%%%%%%%%%%%%%%%%%%%%%%%%%%%%%%%%%%%
\subsection{General linear interaction and $\Lambda$CDM model }
%%%%%%%%%%%%%%%%%%%%%%%%%%%%%%%%%%%%%%%%%%%%%%%%%%%%%

We complete the subject of linear interaction by enlarging the basis elements with a constant, so that, the new base will be $Q_{0}/\D$, $\ro$, $\ro'$, $\ro''$, with $Q_{0}/\D$ a constant. Although, the effective one-fluid model is able to mimic the essential features of a de Sitter scenario, clearly, the introduction of  interactions could produce different alternatives to this scenario.  

We introduce the ``general linear interaction", $(Q_{gL})$,
\be
\n{gl}
Q_{gL}=\frac{Q_{0}}{\D}+Q_L,
\ee
where the constrain (\ref{bc'}) holds for the constants of $Q_L$. A particular type of the general linear combination (\ref{gl}), $Q=C_0+C_1\ro_c+C_2\ro_x$, was analyzed in \cite{Quercellini}. A more general linear interaction can be built including a $f(\eta)$ term to the $Q_{gL}$ (\ref{gl}), see Ref. \cite{luis}. A particular case where the interaction $Q_{gL}\propto f(\eta)$ was examined in \cite{fabris1}. Combining Eqs. (\ref{2}), (\ref{qlf}), (\ref{17}) and (\ref{gl}) the Eq. (\ref{2}) becomes
$$
(1-b_3\D)\ro''+\ga_s^{-1}\left[\ga_s^2+\ga_c\ga_x-(b_1+b_3\ga_s^2)\D\right]\ro'
$$
\be
\n{qgl}
+(\ga_c\ga_x-b_1\D)\ro=Q_{0},
\ee
whose general solution is
\be
\n{sgl}
\ro_{gL}=\La_{eff}+ca^{-3\ga_s}+ba^{-3\ga^+},
\ee
where $\La_{eff}=Q_{0}/(\ga_c\ga_x-b_1\D)>0$ is the effective cosmological constant, induced by the constant term $Q_0/\D$ in the $Q_{gL}$.

At late times the total energy density has the limit $\ro_{gL}\to\La_{eff}$ and the effective equation of state becomes $p\approx -\La_{eff}$. Thus, the effective one-fluid model can be associated with a unified dark sector model whose scale factor interpolates between a power-law phase and a de Sitter stage $H=H_0=\sqrt{3/\La_{eff}}$ being $H_0$ an attractor. 

From Eq. (\ref{31}), the dark energy densities are 
\be
\n{r1gl}
\ro_{cgL}=-\frac{\ga_x\La_{eff}}{\D}+\ro_{c}, \qquad \ro_{xgL}=\frac{\ga_c\La_{eff}}{\D}+\ro_{x},
\ee
where $\ro_c$ and $\ro_x$ are given by Eqs. (\ref{r12'}) and (\ref{r22}). When the total energy density tends to $\La_{eff}$ for $a\to\infty$, the dark matter energy density has the final limit $\ro_{cgL}\to -\ga_x\La_{eff}/\D<0$. To relieve this problem we may assume a phantom equation state for the dark energy, with $\ga_x<0$, so the energy densities (\ref{r1gl}) become positive and the ratio $r_{gL}$ tends to $r_\infty=-\ga_x/\ga_c>0$.

%%%%%%%%%%%%%%%%%%%%%%%%%%%%%%%%%%%%%%%%%%%%%%%%%%%%
\section{Nonlinear interaction}
%%%%%%%%%%%%%%%%%%%%%%%%%%%%%%%%%%%%%%%%%%%%%%%%%%%%

Let us assume that the energy transfer between the dark matter and dark energy components is produced by the following ``nonlinear interaction",
\be
\n{nl}
Q_{nL}=\frac{j_1\ro_c^2+j_2\ro_c\ro_x+j_3\ro_x^2}{\ro}+Q_L+\frac{f(\eta)\ro^\nu}{\D},
\ee
where the $Q_L$ is giving by Eq. (\ref{lcg}), $f(\eta)\ro^\nu$ is a nonlinear atypical term  proportional to a well-behaved function $f(\eta)$, which depends on the scale factor $\eta=\ln{a^3}$, with $\nu$ a constant. 

By using Eq. (\ref{31}) we see that the three terms in the numerator of the nonlinear part of the Eq. (\ref{nl}), become a linear combination of $\ro^2$, $\ro\ro'$ and $\ro'^2$. Rearranging all these terms, we obtain the final form of the $Q_{nL}$ (\ref{nl}),
\be
\n{final}
Q_{nL}=\frac{k_1\ro'^2+\ro(k_2\ro+k_3\ro'+k_4\ro'')+f(\eta)\ro^{\nu+1}}{\ro\D},
\ee
where the $k_i$ are combinations of the constants in Eq. (\ref{nl}). From Eqs. (\ref{2}) and (\ref{final}), we obtain the nonlinear differential equation for $\ro$
$$
\ro\ro''+\frac{\ga_c+\ga_x-k_3}{1-k_4}\ro\ro'-\frac{k_1}{1-k_4}\ro'^2\,\,\,\,\,\,\,\,\,\,\,\,\,\,\,\,\,\,\,\,\,\,\,\,\,\,\,\,\,\,\,
$$
\be
\n{Qnl}
+\frac{\ga_c\ga_x-k_2}{1-k_4}\ro^2=\frac{f}{1-k_4}\ro^{\nu+1}.
\ee
In the particular case where $k_4=1$ and $f=0$, the first and last terms in the RHS of the $Q_{nL}$ (\ref{final}) vanish simultaneously, Eq. (\ref{Qnl}) becomes a homogeneous linear differential equation for $\ro$. However, in other cases, the general solution of the source equation (\ref{Qnl}) will be obtained from a nonlinear superposition of the two basis solutions. 

Renaming the four constants in the Eq. (\ref{Qnl}) by $n_1$, $n_2$, $n_3$ and $n_4$ respectively (from left to right) and changing to the new variable $x=\ro^{(1+n_2)}$ with $n_2\neq -1$ or $k_1+k_4\neq 1$, the Eq. (\ref{Qnl}) turns into    
\be
\n{x''}
x''+n_1x'+n_3(1+n_2)x=n_4(1+n_2)f(\eta), 
\ee
where we have chosen $\nu=-n_2=k_1/(1-k_4)$. If $x_{1h}$ and $x_{2h}$ are the two basis solutions of the homogeneous Eq. (\ref{x''}), then the general solutions of the Eq. (\ref{Qnl}) can be written as a nonlinear superposition of these basis solutions $\ro_{nL}=(c\,x_{1h}+b\,x_{2h}+x_p)^{1/(1+n_2)}$, where $x_p$ is a particular solution of Eq. (\ref{x''}). From now on, the Eq. (\ref{x''}) will substitute the source equation (\ref{2})  for the $Q_{nL}$.

%%%%%%%%%%%%%%%%%%%%%%%%%%%%%%%%%%%%%%%%%%%%%%%%%%%%
\subsection{The homogeneous case}
%%%%%%%%%%%%%%%%%%%%%%%%%%%%%%%%%%%%%%%%%%%%%%%%%%%%

In the $f=0$ case, the  general solution of Eq. (\ref{x''}) is $x_h=c\,a^{3\la^-}+b\,a^{3\la^+}$. Then, coming back to the original variable $\ro_h=x_h^{1/(1+n_2)}$ and using the Eqs. (\ref{31}) and (\ref{pe}), we find 
\be
\n{sg}
\ro_h=\left[c\,a^{3\la^-}+b\,a^{3\la^+}\right]^{1/(1+n_2)},
\ee

\be
\n{r1nl}
\ro_{ch}=-D\left[[\la^-+\ga_x(1+n_2)]\ro+b\,\D\la\,\,\frac{a^{3\la^+}}{\ro^{n_2}}\right],
\ee
\be
\ro_{xh}=D\left[[\la^-+\ga_c(1+n_2)]\ro+b\,\D\la\,\,\frac{a^{3\la^+}}{\ro^{n_2}}\right],
\ee
\be
\n{e1}
p_h=-\left(1+\frac{\la^-}{1+n_2}\right)\ro_h-\frac{b\,\D\la\,\,a^{3\la^+}}{(1+n_2)\ro_h^{n_2}}.
\ee
where $\la^-$, $\la^+$ are the characteristic roots of the Eq. (\ref{x''})
\be
\n{char}
\la^\mp=\frac{-n_1\mp\sqrt{n_1^2-4n_3(1+n_2)}}{2},
\ee
with $D=[(1+n_2)\D]^{-1}$ and $\D\la=\la^+-\la^-$.
Depending on the values of the parameters $n_1$, $n_2$, and $n_3$ the total energy density (\ref{sg}) behaves asymptotically as $\ro_h\to a^{3\la^\pm/(1+n_2)}$ in the limit of large scale factors, meaning that $a\to t^{-2(1+n_2)/3\la^\pm}$. For $c=0$ (+) or $b\,=0$ (-), the model includes the exact power-law expansions $a^\pm=t^{-2(1+n_2)/3\la^\pm}$. In particular, when $(1+n_2)/3\la^\pm>0$, we have a final phantom phase. In the special cases that $\la^\pm=0$ and $\la^\mp<0$, namely $n_1>0$ and $n_3=0$ or $n_1<0$ and $n_3=0$ (see Eq. (\ref{char})), we have a final de Sitter stage with an effective cosmological constant given by the limit $\ro_h\to\La_{eff}=b\,^{1/(1+n_2)}$ or $\ro_h\to\La_{eff}=c^{1/(1+n_2)}$. These models include the modified Chaplygin gas introduced in Ref. \cite{marian} where it was proposed the equation of state $p=A\ro-B/\ro^n$, with $n\geq 1$, and the parameters $A$ and $B$ were constrained to be positive. When both, $\la^+\neq 0$ and $\la^-\neq 0$, the equation of state (\ref{e1}) contains those which characterize various of the variable modified Chaplygin gas models investigated in Refs.  \cite{guo1}-\cite{deb}. 

%%%%%%%%%%%%%%%%%%%%%%%%%%%%%%%%%%%%%%%%%%%%%%%%%%%%
\subsection{Nonlinear examples}
%%%%%%%%%%%%%%%%%%%%%%%%%%%%%%%%%%%%%%%%%%%%%%%%%%%%

Here we examine the case where $Q_{h}$ takes the form
\be
\n{ch}
Q_{h0}=\al\ga_c\frac{\ro_c\ro_x}{\ro},
\ee
with $\al$ constant, see \cite{waga} and references therein. From Eqs. (\ref{Qnl}) and (\ref{ch})  we identify the three coefficients $n_1=(\ga_c+\ga_x)(1+n_2)$, $n_2=\al\ga_c/\D$ and $n_3=\ga_c\ga_x(1+n_2)$. Then, the characteristic roots read
\bn{r+-}
\la^+=-\ga_x(1+n_2),  \qquad  \la^-=-\ga_c(1+n_2).
\ee
Introducing these roots into the general solution (\ref{sg}) and the effective equation of state (\ref{e1}), they reduce to 
\be
\n{scha}
\ro=\left[c\,a^{-3\ga_c(1+n_2)}+b\,a^{-3\ga_x(1+n_2)}\right]^{1/(1+n_2)},
\ee
\be
p=(\ga_c-1)\ro-b\,\D\frac{ a^{-3\ga_x(1+n_2)}}{\ro^{n_2}}.
\ee
This effective equation of state can be identified with those which were used to build variable modified Chaplygin gas models \cite{guo1}-\cite{deb}.  

%%%%%%%%%%%%%%%%%%%%%%%%%%%%%%%%%%%%%%%%%%%%%%%
\subsubsection{Modified Chaplygin gas}
%%%%%%%%%%%%%%%%%%%%%%%%%%%%%%%%%%%%%%%%%%%%%%%

We make an adequate selection of the parameters so that the interaction (\ref{ch}) be focused on a dark energy component described by some kind of vacuum energy density i.e., $\ga_x=0$. Then, $n_3=\ga_c\ga_x(1+n_2)$=0, $\D=\ga_c$, $n_2=\al\ne-1$, $\la^+=0$, and the second term in Eq. (\ref{scha}) becomes constant. Assuming that the dark matter component is nearly pressureless, we may associate it to a barotropic fluid with a free constant parameter $\ga_c=\ga_0\approx 1$. Then the energy density (\ref{scha}) abbreviates to
\be
\n{rocha}
\rho=\left[\frac{B}{\ga_0}\pm\left(\frac{a_0}{a}\right)^{3{\ga_0(1+\alpha)}}\right]^{{1}/{1+\alpha}},
\label{r}\ee
where $B$ is a constant. Hence, by replacing this energy density in Eq. (\ref{pe}), we obtain the equation of state of the one effective fluid
\be
\n{pcha}
p=(\ga_0-1)\rho-\frac{B}{\rho^{\alpha}}\label{p}.
\ee
It characterizes several unified cosmologies implemented with Chaplygin gases  as the generalized, extended, modified and enlarged ones \cite{Kamenshchik}-\cite{Cervantes-Cota}. 

Now, we express the Chaplygin gases with equation of state (\ref{pcha}) as an interacting two-fluid model where the energy exchange is produced by the interaction term (\ref{ch}). To this end, we insert the equation of state (\ref{pcha}) into Eq. (\ref{31}) to find the dark matter and dark energy densities 
\be
\n{r12s}
\ro_{cCh}=\ro-\frac{B}{\ga_0\ro^{\al}},\quad \ro_{xCh}=\frac{B}{\ga_0\ro^{\al}},
\ee
\be
\n{rs}
r_{Ch}=-1+\frac{\ga_0}{B}\,\ro^{\al+1}.
\ee
On the other hand, by combining the Eq. (\ref{rocha}), (\ref{r12s}) and (\ref{rs}), we can express the quantities characterizing the interacting two-fluid model in terms of the scale factor. 

For expanding universes and $\ga_0(1+\al)>0$, the total energy density (\ref{rocha}) $\ro^{\al+1}\to B/\ga_0$ and the ratio has a vanishing limit $r_{Ch}\to 0$. 

%%%%%%%%%%%%%%%%%%%%%%%%%%%%%%%%%%%%%%%%%%%%%
\subsubsection{The $\ga_0=0$ case}
%%%%%%%%%%%%%%%%%%%%%%%%%%%%%%%%%%%%%%%%%%%%%

When $\ga_0=0$, the effective one-fluid model described by the expressions (\ref{rocha}) and (\ref{pcha}) is not valid. However this case can studied from the nonlinear interaction term 
\bn{Qpoly}
Q_{r}=\frac{\al(k_4-1)\ro^{-1}\ro'^2+\ga_c\ga_x\ro+(\ga_c+\ga_x)\ro'+k_4\ro''}{\D}.
\ee
Now, the source equation (\ref{x''}) reduces to $x''=0$ and one finds that the total energy density $\ro=x^{1/(1+\al)}$
\bn{poly}
\ro=\ro_0\left[\pm 1+b\,\ln{\left(\frac{a}{a_0}\right)^3}\right]^{1/(1+\al)},
\ee
has a logarithmic dependence with the scale factor, while the equation of state of the effective one-fluid model is
\be
p=-\ro-\frac{b\,\ro_0}{1+\al}\,\,\left(\frac{\ro_0}{\ro}\right)^{\al}.
\ee
with $\ro_0$ a constant. The last equation of state also can be seen as a generalization of the polytropic equation of state $p=K\ro^{\ga_p}$, where $K$ is a constant and $\ga_p$ is the polytropic index. The scale factor is determined by integrating the Friedmann equation for the source (\ref{poly})  
\bn{spoly}
a=a_0\exp{\frac{1}{3b}\left[\pm 1+\left[\frac{b\sqrt{3\ro_0}(2\al+1)\,t}{2(1+\al)}\right]^{\frac{2(1+\al)}{(2\al+1)}}\right]},\,\,\,\,\,\,\,\,\,\,\,\,\,
\ee
where we have set $t_0=0$. Note that the quantities (\ref{poly})-(\ref{spoly}) are independent of $k_4$, then the interaction term $\ro''$ dos not contribute to the evolution of this unified model. 

%%%%%%%%%%%%%%%%%%%%%%%%%%%%%%%%%%%%%%%%%%%%%%%%%%%%
\subsection{The ``relaxed Chaplygin gas model"}
%%%%%%%%%%%%%%%%%%%%%%%%%%%%%%%%%%%%%%%%%%%%%%%%%%%%

Here we are going to consider the ``inhomogeneous nonlinear interaction" $(Q_i)$,   
\be
\n{final'}
Q_{i}=\frac{k_1\ro^{-1}\ro'^2+k_2\ro+k_3\ro'+k_4\ro''+f(\eta)\ro^{k_1/(1-k_4)}}{\D},
\ee
which has been obtained from the Eq. (\ref{final}) by choosing $\nu=-n_2=k_1/(1-k_4)$. The source equation (\ref{x''}) becomes inhomogeneous and its general solution is given by
\be
\n{sgr}
\ro_{i}=\left[c\,a^{3\la^-}+b\,a^{3\la^+}+x_p\right]^{1/(1+n_2)},
\ee
while the equation of state (\ref{pe}) takes the form
\be
\n{pgr}
p_{i}=-\left[1+\frac{\la^-}{1+n_2}\right]\ro_{i}-\frac{b\,\D\la\,\,a^{3\la^+}+x_p'-\la^-x_p}{(1+n_2)\ro_{i}^{n_2}}.
\ee

(i) For $f=f_0=const$ the particular solution $x_p=n_4f_0/n_3$ is constant and the total energy density (\ref{sgr}) along with the effective equation of state (\ref{pgr}) describe a ``double unified model", in a sense that initially the universe is dominated by the ``two" terms inside the bracket of $\ro_{i}\approx (c\,a^{3\la^-}+b\,a^{3\la^+})^{1/(1+n_2)}$. These terms can be seen as a ``nonlinear mixture of two fluids". But at late times the universe is dominated by a vacuum energy $\ro_{i}\approx x_p^{1/(1+n_2)}$ and has a de Sitter expansion. So, this ${(i)}$ case includes various generalizations of the modified Chaplygin gas model investigated in Refs. \cite{guo1}-\cite{deb}. When one of the constants $c$ or $b\,$ vanishes the effective equation of state (\ref{pgr}) turns in Eq. (\ref{pcha}) and the double unified model produces different versions of the Chaplygin gas \cite{marian}-\cite{Cervantes-Cota}.

(ii) For $f(\eta)\ne const$ and $\eta=\ln{a}$, the particular solution $x_p$ and consequently the numerator of the last term in the RHS of the effective equation of state (\ref{pgr}) become arbitrary functions of the scale factor. Then, a fluid obeying the Eq. (\ref{pgr}) defines a ``relaxed Chaplygin gas model".

%%%%%%%%%%%%%%%%%%%%%%%%%%%%%%%%%%
\section{Conclusions}
%%%%%%%%%%%%%%%%%%%%%%%%%%%%%%%%%%

We have investigated interacting dark sector models with energy transfer and introduced  an effective one-fluid description with an effective equation of state. The two coupled equations describing the interacting model have been combined to obtain the fundamental second order differential equation for the total energy density, ``the source equation". We have assumed a separable interaction $Q=\ro Q(\ga,\ga')$, which includes a large set of  cases investigated in the literature, and examined the stability condition for the power-law solutions. 

We have presented the ``linear interaction" $Q_L=c_1\ro_c+c_2\ro_x+c_3\ro_c'+c_4\ro_x'+c_5\ro+c_6\ro'+c_7\ro''$, and reduced it to a linear combination of the basis elements $\ro$, $\ro'$ and $\ro''$, so that  $Q_L=b_1\ro+b_2\ro'+b_3\ro''$. As far as we know, this interaction has not been investigated in the literature. We have found the stationary solution for the effective barotropic index, $\ga_s$ and $\ga^+$, and imposed the stability condition, being the power-law solution $a=t^{2/3\ga_s}$ an attractor. Interestingly, the existence of the $\ga_s$ solution is linked to the fulfillment of the requirement $Q_L(\ga_s)<0$ on the attractor, indicating that the energy is being transfered from the dark energy to the dark matter. We have considered several particular examples and observed that $Q_L\propto\ro_x$, $Q_L\propto\ro$ and $Q_L\propto\ro'$ are satisfactory couplings because in each case, the ratio $r_s=\ro_{cs}/\ro_{xs}$ is an attractor. These simple models may alleviate the problem of the coincidence. Although, a coupling proportional to the dark matter energy density does not lead to stable solutions however, it can work when it is combined with the remaining terms of $Q_L$. Finally, we have obtained the exact scale factor and the effective barotropic index in implicit form. 

We have generalized the above coupling by adding a constant to the $Q_L$ and introduced the ``general linear interaction", $Q_{gL}=\La_{eff}(\ga_c\ga_x-b_1\D)/\D+Q_L$. These models lead to several alternatives to the $\La$CDM model.

We have presented a class on ``nonlinear interaction" $Q_{nL}=[k_1\ro^{-1}\ro'^2+k_2\ro+k_3\ro'+k_4\ro''+f(\eta)\ro^{k_1/(1-k_4)}]/\D$. Although the source equation becomes a nonlinear differential equation, we have linearized and solved it. In general, we have found that the equation of state of the effective one-fluid model depends explicitly on the scale factor and the universe evolves to a power-law scenario for large cosmological times. However for interactions having the form $Q_{nL}=\al\ga_c\ro_c\ro_x/\ro$ with $\al$ a constant, we have shown that the effective equation of state becomes that of the Chaplygin gas when the dark energy component is described by some kind of vacuum energy density i.e., $\ga_x=0$. These unified model has been expressed as an interacting one. Also, we have examined the particular nonlinear interaction leading to the source equation $x''=0$. Here we have shown that the effective equation of state can be seen as a generalization of the polytropic equation of state.

Generically, when there are no restrictions on $Q_{nL}$, the equation of state of the effective one-fluid model defines what we have called the ``relaxed Chaplygin gas model". It contains various generalizations of the Chaplygin gas, including the variable modified Chaplygin gas model with equation of state $p=A\ro+B(a)/\ro^\al$.

%%%%%%%%%%%%%%%%%%%%%%%%%%%%%%%%%%%%%%%%%%%%%%%%%%
\acknowledgments
%%%%%%%%%%%%%%%%%%%%%%%%%%%%%%%%%%%%%%%%%%%%%%%%%

The author thanks the financial support of the I-CosmoSul's organizers to participate in the event,
the University of Buenos Aires for the partial support of this work under Project X044 and the Consejo Nacional de Investigaciones Cient\'\i ficas y T\'ecnicas under Project PIP 114-200801-00328.

\end{document}